# Optical absorption preceding resonant double photoionization of aromatic hydrocarbons

D. L. Huber[a]

Physics Department, University of Wisconsin-Madison, Madison, WI 53706, USA

**Abstract**

We analyze resonances in the double photoionization of a variety of aromatic hydrocarbons. The resonances reflect the breakup of quasi-bound electron pairs. The basic premise of this paper is that there is a direct connection between the quasi-bound pairs and resonant peaks in the optical absorption that are associated with doubly occupied sites on the perimeter and inside the perimeter of the molecule. The optical absorption leading to the high-energy resonance (approximately 40 eV), calculated from a many-site one-dimensional Hubbard model, has a peak at $U$, the electrostatic interaction energy for two electrons with antiparallel spins on the same carbon atom. In the model, there are also two satellites whose separation from the main resonance is approximately $\pm 10$ eV suggesting that unresolved satellite structure may be contributing to the linewidth of the resonant peak. The low energy resonances (approximately 10 eV) involve carbon atoms located inside the perimeter, a configuration present only in pyrene and coronene (among the hydrocarbons studied). In the case of pyrene, which has two carbon atoms inside the perimeter, we employ a two-site Hubbard model to characterize the absorption leading to the quasi-bound state. A brief analysis of the double photoionization resonance of the heterocyclic inorganic molecule 1,3,5-triazine presented. We also discuss recent results for the double photoionization of the cyclic inorganic molecule tribromoborazine and the organic molecules furan, pyrrole, selenophene, and thiophene where the 2+ ion concentration varies linearly with the difference between the photon energy and the threshold energy. A theory for the linear behavior is outlined.

[a]email: dhuber@wisc.edu

## 1. Introduction

In a recent series of papers [1-5], R. Wehlitz and his colleagues reported the results of detailed studies of two-electron photoionization from a variety of cyclic and polycyclic organic molecules. In the case of (deuterated) benzene and other aromatic compounds comprised of carbon and hydrogen, they found a peak in the ratio of doubly ionized to singly ionized parent ions that is $\approx 40$ eV above threshold. They attributed this peak to the formation of a mobile, two-electron quasi-bound state above the LUMO band. In their interpretation, the photoionization is a two-step process in which a quasi-bound state is created and subsequently decays into two free electrons. In this picture, the question arises as to the nature of the quasi-bound state. Originally it was suggested that the bound state might be a Cooper pair, as found in superconducting materials, but there are no detailed theoretical studies that support this picture. In an earlier paper [6], we proposed an alternative mechanism involving the formation of Coulomb pairs. 'Coulomb pair' refers to the quasi-bound state of two electrons in a one-dimensional array with periodic boundary conditions [7,8]. In the present context, the one-dimensional periodic array is associated with the carbon atoms that on perimeter of the molecule. In the Coulomb pairing interpretation, the resonance peak occurs at approximately the energy of the electrostatic interaction between π electrons with antiparallel spins occupying the same carbon cite. In [6], we calculated the bound state energy using atomic orbitals and obtained a rough estimate of 18.5 eV.

## 2. High energy resonances

The starting point in our analysis of the two-electron resonance is the existence of a two-electron quasi-bound state. The basic premise of this paper is that there is a direct connection between the quasi-bound pairs and the 40 eV resonances in the optical absorption. We address the problem using the one-dimensional Hubbard model to calculate the optical absorption [9-11]. The Hubbard model was originally developed in a study of magnetism and related phenomena in solids. Since the model is based on the tight-binding approximation for the electronic states, it has found application in organic chemistry for molecules whose low-lying electronic states are molecular orbitals, the analog of tight-binding wave functions in crystals. In one dimension, the Hubbard model is directly connected to the Coulomb pairing state [9]. It is applicable to benzene where all carbon atoms are equivalent; in the case of the polycyclic hydrocarbons, we neglect differences in perimeter sites and apply the model to all the carbon atoms on the perimeter of the molecule. In the model, the Hamiltonian for $N$ equivalent sites is written as

$$H = -t \sum_{j=1}^{N} \sum_{\sigma} (c^*_{j,\sigma} c_{j+1,\sigma} + c^*_{j+1,\sigma} c_{j\sigma}) + U \sum_{j=1}^{N} n_{j\uparrow} n_{j\downarrow} \qquad (1)$$

in second quantized form [9,10]. In this equation, $c_{j,\sigma}$ denotes the electron annihilation operator for site $j$ and spin $\sigma$, and $c^*_{j,\sigma}$ is the corresponding creation operator. When the boundary conditions are periodic, we have $c_{N+1,\sigma} \equiv c_{1,\sigma}$ etc. The parameter $t$ is the transfer integral connecting nearest-neighbor sites while the parameter $U$ is the energy of the electrostatic interaction between two electrons with antiparallel spins (thus satisfying the exclusion principle) occupying the same site. Multiplying it are the number operators for the up and down spins $n_{j,\uparrow}$ and $n_{j,\downarrow}$. When $U = 0$, the Hubbard Hamiltonian reduces to the familiar Hückel model for molecular orbitals where the band energy is of the form $-2t \cos(k)$ whereas when $U$ is finite there can also be charge transfer transitions to doubly occupied sites. The Hubbard Hamiltonian also has points in common with the semi-empirical approach to π-conjugated organic molecules developed by Pariser and Parr and by Pople in the 1950s.

With the high-energy resonances, we are in the limit where $U \gg t$. In analyzing the behavior in this limit, we make use of the results reported in [10] (Section V) for the optical absorption of the one-dimensional Hubbard model with periodic boundary conditions. Figure 10 of that reference shows numerical results for the absorption in a twelve-site array with $t = 1$, $U/t = 6, 12,$ and $24$ and half filling. It is evident that the absorption has a dominant peak at resonance energy $U$. In addition, there are major satellite peaks at approximately $U \pm 3$ corresponding to a satellite to main peak separation of 3t [10]. Assuming $t = 2.54$ eV, a value appropriate for benzene [11], this result suggests that the satellite structure may be making a significant contribution to the observed resonance linewidth which is approximately 10 eV (halfwidth at half height, low energy side). It should also be noted that with the Hubbard model, in the limit $N \to \infty$, the optical absorption for $U/t \gg 1$ is limited to the region $U - 4t < E < U + 4t$ [12,13].

## 3. Low energy resonances: pyrene and coronene

Of the aromatic hydrocarbons investigated so far, benzene, naphthalene, anthracene, pentacene, phenanthrene, azulene, pyrene and coronene, only pyrene and coronene display a low energy peak in the ionization ratio [4]. In both cases, the peak is approximately 10 eV above threshold. From Fig. 1, it is evident that pyrene and coronene have carbon atoms inside their perimeters. Of the molecules cited above, they are the only ones that have interior carbon atoms. We propose that the quasi-bound states leading to the 10 eV photoionization peaks are associated with the interior carbon atoms. We focus on pyrene, where the Hubbard model analysis is relatively simple, as it involves only two carbon atoms.

In [12], the authors present results for the optical properties of the dimerized Hubbard model with alternating transfer integrals $t(1 \pm \delta)$. When $\delta = 1$, The system reduces to an array of isolated pairs. The eigenstates of the Hubbard Hamiltonian for the two-site model that are involved in the optical absorption are the ground state with energy

$$E_0^- = U_{int}/2 - \left[(U_{int}/2)^2 + 4t_{int}^2\right]^{1/2} \quad (2)$$

and the doubly occupied excited state with energy $U_{int}$ (Eqs. 15,19 and 20 in [12]), We identify the resonance energy, $RE$, with the difference between the two energy levels:

$$RE = U_{int}/2 + \left[(U_{int}/2)^2 + 4t_{int}^2\right]^{1/2} \quad (3)$$

In applying Eq. (3) to pyrene, we assume $t_{int} = 2.54$ eV [11] leading to a value of $U_{int}$ of approximately 7.4 eV, which is well within the HOMO-LUMO manifold. The shape of the low-energy resonance in pyrene is approximately triangular and similar to the shape of the corresponding resonance in coronene suggesting that the two-site model is appropriate there, at least as a first approximation. It differs, however, from the shape of the high-energy resonance involving the perimeter atoms. We attribute this difference to the fact that in the two-site model, the low-energy resonance reflects a transition between isolated levels whereas multiple eigenstates are involved in the high energy transition [12,13].

## 4. Summary and discussion

In this paper, we have presented a semi-quantitative theory for the resonances in the two-electron photoionization of aromatic hydrocarbons. The theory is based on the one-dimensional Hubbard model where the resonant absorption of the incoming photon leads to a two-electron quasi-bound state or Coulomb pair. In our analysis, we made extensive use of the numerical and analytical results reported in [10,12]. All the aromatic hydrocarbon molecules that have been studied so far have high-energy resonances at $\approx 40$ eV that we associate with carbon atoms on the perimeter of the molecule. Unique among the molecules studied, coronene and pyrene also have low energy resonances ($\approx 10$ eV) that we attribute to quasi-bound states of $\pi$ electrons on carbon atoms inside the perimeter. The difference in the energies of the high and low energy resonances reflects the difference in the magnitude of the parameters $U$ and $U_{int}$.

The high-energy resonances have energies much greater than the transfer integral. In this limit, the resonance peaks are centered at $U$. In the $N$-site Hubbard model ($N \gg 2$), when $U/t \gg 1$ there is substructure in the form of satellites at $U \pm 4t$ leading to a half width of $4t$ [12,13]. For the perimeter sites, the half widths of the high-energy peaks are $\approx 10$ eV corresponding to a transfer integral $\approx 2.5$ eV. Wehlitz et al. [1,5] have stressed the point that the high-energy peak is at an energy that matches the kinetic energy of a two-electron quasiparticle with a de Broglie wavelength equal to the $C$-$C$ separation. Their finding is consistent with the identification of the doubly excited site in the Hubbard model as the analog of a de Broglie-like wave packet.

We also call attention to a series of heterocyclic molecules having benzene-like structures: pyridine with one nitrogen atom and five carbon atoms, pyrimidine, with two nitrogen atoms separated by a carbon atom, and 1,3,5-triazine, which has a benzene structure with six alternating carbon and nitrogen atoms  Recent measurements of the double photoionization by Hartman and Wehlitz [14] show that the pyridine and pyrimidine distributions in the ratio of doubly to singly charged ions are greatly reduced in comparison with benzene when scaled according to their knock-out contributions. In the case of 1,3,5-triazine the scaled peak in the ratio is shifted to 50 eV but has approximately the same height and width as in benzene. The Hubbard theory provides a simple interpretation of the triazine results. The symmetry is captured in a six-site two-sublattice model with periodic boundary conditions ($U_i = U_N$, $i = 1, 3, 5$; $U_i = U_C$, $i = 2, 4, 6$). As with the aromatic hydrocarbons, when $U \gg t$, the double ionization ratio has a central peak at $U$ with subsidiary structure associated with the transfer integral. In [13], it is pointed out that the energy of the peak is approximately the kinetic energy of a particle with mass $2m_e$ and a de Broglie wavelength equal to the $C$-$N$ separation.

It should also be noted that the cyclic property alone does not lead to a threshold resonance. It is also necessary to have charge transfer to form the Coulomb pair. This is seen in recent investigations of the double photoionization of the cyclic inorganic molecule tribromoborazine (BrBNH)$_3$, which is an analog of 1,3,5-triazine. In this case, there is no resonance and the concentration of 2+ ions increases linearly above the threshold [15]. Similar results have been obtained for the organic molecules furan, pyrrole, selenophene and thiophene [5,16].

Although a detailed theory of the linear increase is not available, there is a qualitative interpretation where the absorption of the incident photon gives rise to an excited state that decays into an excited 2+ ionic state and two free photoelectrons [6]. The photoelectrons' total kinetic energy is equal to the difference between the photon energy and the threshold energy. Under steady-state conditions, the 2+ population is determined by the two-electron transition rate $W$. Since the photoelectrons are uncorrelated, $W$ is proportional to the product of the individual final-state energy densities of the two electrons which vary as the 1/2 power of their kinetic energy [17]. The linear behavior arises when the kinetic energies of the two electrons, $E_1$ and $E_2$, are proportional to $\Delta$, i.e. $E_1 = f\Delta$ and $E_2 = (1-f)\Delta$. The corresponding transition rate is then expressed as

$$W = C[f(1-f)]^{1/2}\Delta \qquad (4)$$

where $C$ is independent of $\Delta$. Because of the relatively small photon momentum, the two photoelectrons move in nearly anti-parallel directions with kinetic energies approximately equal to $\Delta/2$, behavior similar to what is predicted for the double photoionization of helium in the high-energy limit [18].

An important issue connected with our analysis that is not addressed by the Hubbard model is the lifetime of the Coulomb pair. In the case of the two-site analysis, the pair state is an isolated eigenstate and has an infinite lifetime (neglecting radiative decay). The critical issue here is that the Hubbard model neglects interactions with core electrons and other valence electrons. In [8], a rigorous proof is given that two electrons in a one-dimensional potential with periodic boundary conditions can form a bound state with infinite lifetime. In the case of the cyclic hydrocarbons, the lifetime of the pair appears infinite only when one approximates the interactions with the other electrons by a periodic potential. In view of the importance of the lifetime, it would be worthwhile to see if it can be determined from the double-photoionization response to pulsed optical excitation. An equally important theoretical issue is the magnitude of the Coulomb pairing energy. The calculation of the pairing energy reported in [6] is based on a primitive approach. A detailed quantum chemistry calculation is needed.

**Acknowledgment**

We thank Ralf Wehlitz and Chun C. Lin for helpful comments.

**Pyrene**

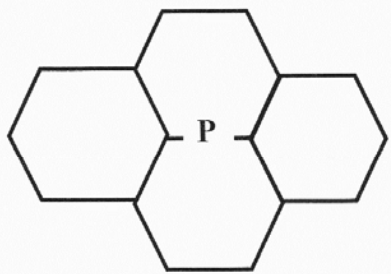

**Coronene**

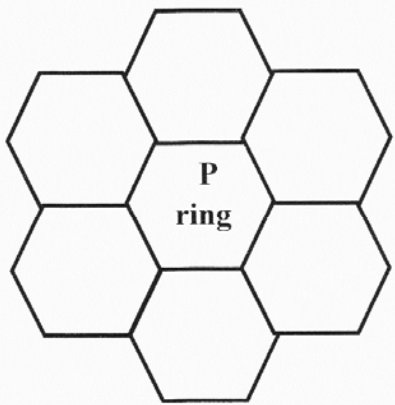

Fig. 1. Pyrene and coronene molecular structures. In the case of pyrene, P identifies the bond between the two carbon atoms associated with the low-energy pairing, whereas with coronene, P is at the center of the ring of six carbon atoms that are linked to the low-energy pairing.